\documentclass[12pt, preprint]{aastex}
\usepackage{graphics}

\shorttitle{PAH Emission Deficit In Low-Metallicity Galaxies}
\shortauthors{O'Halloran et al.}

\begin{document}

\title{The PAH Emission Deficit In Low-Metallicity Galaxies - A Spitzer View}

\author{B. O'Halloran\altaffilmark{1}, S. Satyapal\altaffilmark{1,2} and R. P. Dudik\altaffilmark{1}}
\email{boh@physics.gmu.edu}

\altaffiltext{1}{Dept. of Physics \& Astronomy, George Mason
University, Fairfax, VA, 22030, USA} \altaffiltext{2}{Presidential
Early Career Award Scientist}

\begin{abstract}

Archival observations of 18 starburst galaxies that span a wide
range in metallicity reveal for the first time a correlation between
the ratio of emission line fluxes of [FeII] at 26 $\mu$m and [NeII]
at 12.8 $\mu$m and the 7.7 $\mu$m PAH strength, with the
[FeII]/[NeII] flux ratio decreasing with increasing PAH strength. We
also find a strong correlation between the [FeII]/[NeII] flux ratio
and the host galaxy metallicity, with the flux ratio decreasing with
increasing metallicity.

Since [FeII] emission  has been  linked primarily  to supernova
shocks, we attribute the  high [FeII]/[NeII] ratios in
low-metallicity galaxies to enhanced supernova activity. We consider
this to be a dominant mechanism for  PAH  destruction,  rather than
grain destruction in photoionized regions surrounding young massive
stars. We also consider whether  the   extreme  youth  of  the
low-metallicity galaxies  is responsible for the lack of PAH
emission.

\end{abstract}


\section{Introduction}

Recent  surveys  in  the  UV and  IR
\citep{gal05,eng05,hogg05,mad05,ros05} have highlighted  the
importance of  low-metallicity  galaxies for  the investigation of
vigourous episodes of star formation.  In particular, nearby  blue
compact dwarf  (BCD)  galaxies  provide evidence  that isolated,
small, low-metallicity  galaxies may experience  high star formation
levels in the present epoch \citep{hop02}.  Since the advent of {\it
ISO} it  has been  known that very  low metallicity  ($\it Z$ $\ll$
1) galaxies  show very little in the  way of polycyclic aromatic
hydrocarbonate   (PAH) \citep{pug89}    emission   in   the   mid-IR
\citep{mad00,stu00}.  PAH emission  is thought to be linked  to star
forming activity \citep{rou01}  - UV photons  from massive young
stars excite the PAH carriers, and therefore trace the extent of
star formation.

Low PAH  luminosity in strongly  star forming galaxies is  contrary
to what has  been seen previously by  {\it ISO} surveys  - in
starbursts, ubiquitous  strong PAH  emission was  detected (see
review by Genzel \& Cesarsky 2000). However from  {\it  ISO}  data
it has been tentatively noted that as  the metallicity of the
galaxy increases, so too does the strength of the PAH  emission
\citep{stu00,haas02}. This begs the question - what causes the PAH
emission deficiency in low-metallicity galaxies?  Is it a case of
PAH carriers being destroyed or is the PAH deficiency intrinsic?

Hirashita et al. (2002) note that dust destruction is the dominant
ISM process when the metallicity of  a low metallicity galaxy such
as blue compact   dwarfs    (BCDs)   reaches    12   +   log (O/H)
$\leq$ $\sim$8. Furthermore,  recent work  by \citet{gal05} suggest
that the small size of ISM grains emitting  in the mid and far-IR
may be  due to destruction by shock waves from supernovae (SN) in
these galaxies.  In  low-metallicity galaxies, star formation  is
dominated by  massive stars (M$_{init}$ $\geq$ 35M$_{\odot}$) and
high  SN rates are expected in galaxies experiencing  very recent
massive  star formation \citep{mas99}. An extremely  high SN  rate
and/or extremely  energetic  SN  (with up to  10$^{52}$  ergs  being
released  into the ISM \citep{gal98})  from very massive stars could
very well be responsible for the  lack of PAH in low-metallicity
galaxies. In this scenario, strong SN shocks propagating into the
ISM may be the culprit responsible for the PAH deficiency.

In  order to investigate  this question,  a tracer  of SN  activity
is required.    The    mid-IR     [FeII]    lines    offer    an
ideal extinction-insensitive tool to probe  the SN activity in
actively star forming galaxies.  Iron is  an abundant refractory
element; however it is highly  depleted from the gas  phase of the
interstellar medium in galaxies  as  a  result   of  condensation
onto  dust  grains  (e.g., de Boer, Jura \& Shull 1987).  Forbidden
line emission from low ionization states of Fe is  greatly enhanced
behind hydrodynamic shock fronts  where grain processing  can result
in near  solar gas phase Fe  abundances.  The near-infrared [FeII]
lines at 1.257 $\mu$m and 1.644 $\mu$m have been widely    used to
trace    the SN content    in    galaxies (e.g., Greenhouse et al.
1991, 1997; Calzetti 1997). Likewise, the $\it a^{6}D$(7/2-9/2)
transition of [FeII] near 26 $\mu$m can also be used as a relatively
pure tracer of SN activity. The low excitation potential
corresponding to the transition  (550
K) may in fact imply that
emission from  the mid-IR lines  is longer lasting than emission
from  the higher excitation potential near-IR lines, which are
thought  to persist only up  to 10$^{4}$ years \citep{oli89} after a
SN  explosion. This  feature, coupled with emission from [NeII] at
12.8 $\mu$m which traces emission from very massive young stars,
could be used to determine the SN shock emission  relative to the
present ionizing photon production rate. [NeII]  is  a robust
indicator   of  the ionizing photon rate  in dust-enshrouded
starburst  galaxies.  For galaxies dominated by typical stellar UV
fields, a simple linear proportionality between the [NeII]
luminosity and the  Lyman continuum luminosity is expected from
photoionization  models \citep{tho00}. Measurements  of the
[FeII]/[NeII] emission line ratio in starburst galaxies  could thus
provide a sensitive measure of the supernova content of galaxies
\citep{gre91,gre96,moo88} and hence an  estimate of the  degree of
grain  processing  and  destruction. Furthermore, an increased SN
rate or shock  wave intensity, indicated by a higher [FeII]/[NeII]
 would be  highly suggestive that in low metallicity galaxies destruction by SN shocks
is responsible for the absence of PAHs.

Given the availability of high spatial and spectral resolution data
in the mid-IR from  {\it Spitzer} of nuclear regions  of starbursts
(where massive  stars are  located), we  can search  for [FeII]
emission from nearby  starbursts, and  compare  the [FeII]/[NeII]
ratio as a function  the strength of the PAH emission to  see if any
correlation exists. Such a diagnostic, using  data from  galaxies of
known metallicity, will us to  determine whether  SN driven shocks
play an important role in the PAH  emission deficit in
low-metallicity galaxies.

\section{Observations and data analysis}

A total of 18 starburst  galaxies of known metallicity from optical
studies (see references  in Table  1  for sources)  were selected
from the {\it Spitzer} archive to form the core sample. The full
list of targets is given in Table 1, and are listed in terms of
increasing metallicity. These galaxies range in metallicity from
extremely low (such as I Zw 18, with $ \it Z/Z_{\odot}$=  1/50) to
super-solar metallicity ($\geq \sim 1~\it Z_{\odot}$) galaxies such
as M 82 and NGC 7714. To differentiate between low and
high-metallicity galaxies,  we use a metallicity (12 + log (O/H))
cut-off value of 8.85 - just less than solar, with the majority of
the low-metallicity galaxies having values less than 8.3. The
galaxies range widely in morphology from blue compact dwarfs such as
I Zw 18 to spirals such as NGC 7714 and IC 342. None of the galaxies
in our sample are known to harbour AGN. This is important, as PAH
destruction can occur close to an AGN due to the hard ionization
environment \citep{stu00}. In addition [FeII] emission can be
elevated in galaxies harbouring AGN (e.g. Sturm et al. 2000). One
possible exception is NGC 7714, which is optically classified  as  a
LINER \citep{tho02}. However, the lack of  [NeV] emission at  14 and
24 $\mu$m and  the absence of any evidence for an obscured AGN by
recent Chandra imaging \citep{bran04,smi05} strongly suggests that
NGC  7714 is a pure starburst and it is therefore included in our
sample.  Most of the targets are nearby, with only Mrk 25 (42.1
Mpc), Mrk  930 (77.3 Mpc) and UM 448 (78.4 Mpc) greater than  40 Mpc
away  (assuming a value of $  \it H_{o}$ = 71 km/s/Mpc), allowing
detections of the comparatively weak [FeII] emission.

We extracted low  and high resolution archival spectral  data from
the Short-Low  (SL) (5.2  -  14.5  $\mu$m), Short-High  (SH)  (9.9 -
19.6 $\mu$m) and  Long-High (LH) (18.7 -  37.2 $\mu$m) modules  of
the {\it Spitzer} Infrared Spectrograph (IRS).   The datasets were
derived from a number of {\it Spitzer}  Legacy and Guaranteed  Time
Observation  programs released to the  {\it Spitzer} Data  Archive,
and  consisted of  either spectral  mapping  or staring
observations. We obtained fluxes for the nuclear positions only from
the mapping observations.  All the staring observations were
centered on the galaxy's nucleus.

The data were preprocessed by the {\it Spitzer} Science Center (SSC)
data reduction  pipeline version  12.0 \footnote[2]{{\it Spitzer}
Observers Manual, URL:
http://ssc.spitzer.caltech.edu/documents/som/} before being
downloaded. Further processing was done within the IDL based IRS
data reduction and analysis package {\it SMART},  v  5.5.2
\citep{hig04}. The high-resolution spectra  were extracted  using
the full-aperture extraction method  along  the diffraction orders.
For the low-resolution spectra, the spectra were extracted  using
the interactive  column extraction  option which  is similar to the
method used  in the SSC pipeline, and allowed for definition of a
source as well as a manual definition of the background.  For both
high and low resolution spectra, the ends  of each order  where the
noise increases significantly were manually clipped, as were hot
pixels.

Fluxes were extracted for the emission lines [NeII] at 12.8 $\mu$m
and [FeII]  at 26 $\mu$m  using the  SMART interactive  line-fitting
tool, along with the 7.7 $\mu$m PAH  feature. The 7.7 $\mu$m
PAH strength was determined according to the prescription adopted in
a number of previous studies \citep{rig99,fors04}. Here, the
continuum was determined at the center of the 7.7 $\mu$m feature,
with the continuum determined using a first order linear fit to the
5.9 and 9.5 $\mu$m bandpass. The PAH strength was then calculated as
the ratio of the 7.7 $\mu$m feature peak intensity to the underlying
continuum. The silicate absorption feature at 9.7 $\mu$m was not
prominent in any of our spectra.  We therefore do not consider it to
significantly affect the strength of the 7.7 $\mu$m PAH feature.

The  use of 3 separate apertures for the  SL, SH and LH modules
raises  the issue of aperture effects  on  the extracted  fluxes,
and by extension, any derived correlations. Nuclear star formation,
responsible for the bulk of the PAH emission and mid-IR continuum in
the vast majority of galaxies, originates from the central few
hundred parsecs (e.g. Surace \& Sanders 1999; Scoville et  al. 2000,
Satyapal  et al. 2005). However given the different morphologies and
physical scales within the sample, it is quite plausible that for
some of our sample at least, IR emission and associated star
formation extends further than the central few hundred parsecs and
thus a larger proportion of the MIR and FIR emission may fall
outside the IRS aperture beams.

In order to determine if aperture corrections are required, we used
MIPS 70 $\mu$m images to trace the spatial extent of the FIR
emission. If the FIR was not contained within the Long-Hi slit
(covering roughly the central kiloparsec), we considered the source
to be extended, otherwise the source was considered compact. For 66
\% of the sample (all of the low metallicity and one of the high
metallicity galaxies), the FIR emission is fully contained within
the IRS aperture beams and for these galaxies aperture corrections
are not likely to be significant.  For the remaining galaxies, a
quick inspection of the MIPS 70 micron images reveal that the FIR is
more extended than the IRS aperture. In Sect. 3.1.1, we discuss the
possible impact of the resulting aperture variations on our
results.

\section{Results and Discussion}

\subsection{Behavior of the [FeII]/[NeII] ratio}

\subsubsection{Does the [FeII]/[NeII] ratio vary with PAH strength?}

In order to investigate how the  strength of SN shocks relative to
the massive star population varies with  the strength of the PAH
emission, we plot  the [FeII]/[NeII] emission  line ratio versus the
7.7 $\mu$m PAH  peak to  continuum ratio  for each  galaxy  in
Fig.~1.   It   is  immediately  apparent that  a  strong correlation
exists between  the  [FeII]/[NeII] ratio and  the PAH strength  - as
the PAH  strength increases, the [FeII]/[NeII] ratio drops.
Interestingly, a clear divide exists between low and high
metallicity galaxies on the plot. Employing a Spearman rank
correlation analysis \citep{ken76} to assess the statistical
significance of this trend yields a correlation coefficient
(r$_{s}$) of -0.870 between [FeII]/[NeII] and the PAH strength with
a probability of chance correlation (P$_{s}$)  of 2.73 x 10$^{-6}$,
indicating a significant anti-correlation. The Spearman rank
correlation technique has  the advantage of being non-parametric,
robust to outliers and does  not presuppose a linear relation.

Since the  PAH strength can also  be dependent on the  strength of the
mid-IR  dust continuum,  we can  attempt  to decouple  this effect  by
plotting the  [FeII]/[NeII] ratio versus  the ratio of the  7.7 $\mu$m
PAH feature  luminosity and  the total IR  luminosity (Fig.  2), which
also  includes emission from  much larger  dust grains.   Although the
IRAS  aperture  is  larger than  those  on  IRS,  it has  been  widely
demonstrated that the 12 to 100 $\mu$m fluxes in star forming galaxies
are dominated by emission from  the nuclear regions of the galaxy that
also give rise to the PAH emission \citep{ima03,rod03}.  A preliminary
analysis of the 70 $\mu$m MIPS  images of our sample of galaxies shows
that for  6 of the  7 high metallicity  starbursts included in  Fig. 2
(M82, NGC  7714, NGC 3049, NGC 1482,  NGC 253, and NGC  2903), the FIR
emission is more extended than  the IRS aperture through which the PAH
emission is measured. Since  these aperture variations can potentially
result  in  systematic errors  in  the  PAH/FIR  flux ratio,  we  have
distinguished extended  sources from compact sources in  Fig. 2. Using
the Spearman  rank correlation technique  again, a similar  but weaker
(r$_{s}$  =  -0.738;  P$_{s}$  =  4.73 x  10$^{-4}$),  correlation  is
found. Both  plots confirm that  a strong relationship  exists between
the PAH strength and the strength of the [FeII]/[NeII] ratio. The MIPS
images of  our sample  of galaxies  will be discussed  in a  follow up
work.

\subsubsection{Does the [FeII]/[NeII] ratio vary with metallicity?}

Since it has  been previously suggested from {\it Spitzer} imaging
and ISOCAM-CVF spectra \citep{eng05,mad05,ros05} that PAH emission
also varies with metallicity,  we can also explore whether the
[FeII]/[NeII] ratio varies  in  a similar fashion, given the link
between PAH emission and the [FeII]/[NeII] emission as seen in Fig.
1. In Fig.~3, we see a strong anti-correlation (r$_{s}$ = -0.856;
P$_{s}$ = 5.96 x 10$^{-6}$), with the [FeII]/[NeII] ratio dropping
with increasing metallicity. Again as in Fig. 1, we see a clear
separation in the [FeII]/[NeII] ratio between galaxies of low and
high metallicity.

We must state that the correlations we see here are derived from
compact starburst regions, and we must be careful in extrapolating
this result to suggest that there is a global lack of PAHs in
low-metallicity galaxies. Quiescent dwarf galaxies, which are not
currently undergoing strong star formation or numerous SN events,
may exhibit stronger PAH emission than that seen in the starburst
dwarf galaxies in our sample. For quiescent dwarfs, diffuse star
formation provides a mechanism for the excitation of the PAH
carriers, excited through extended low level star formation and by
lower mass stars from earlier bursts. In support of this, using
fluxes from IRAC and MIPS imaging \citet{eng05} note a large scatter
of PAH emission to 24 $\mu$m continuum ratios for a number of dwarf
galaxies, suggesting that a contribution is made to the overall PAH
emission from quiescent regions outside the nuclear starburst
regions, and would enhance PAH emission even within low metallicity
galaxies.

\subsection{What causes the lack of PAH emission in low-metallicity galaxies?}

\subsubsection{UV radiation or SN shocks?}

UV radiation from  the  massive  O  star population  may  play an
important role  in  PAH  destruction in low metallicity galaxies due
to the hard ionizing radiation field surrounding these  stars
\citep{pla02}.  In Fig.~4, we plot the [NeIII] 15.5 $\mu$m /[NeII]
12.8 $\mu$m ratio versus the PAH strength for our sample.  The
[NeIII]/[NeII] ratio is a robust extinction and
abundance-insensitive indicator of  the hardness of the radiation
field surrounding massive young stars \citep{tho00}. Although Fig.~4
shows a significant correlation (r$_{s}$ = -0.612  ; P$_{s}$= 6.88 x
10$^{-3}$), there is considerable scatter and  the correlation is
weaker  than  the correlation displayed  in Fig.~1. This correlation
confirms a similar diagnostic presented in \citet{mad05} for a range
of objects of varying morphology and spatial scale that include HII
regions, spiral, starburst and dwarf galaxies.

The strength of the correlation in Fig.~4 in relation to that in
Fig.~1 suggests that PAH destruction by SN-driven shocks likely
plays a more pronounced role than grain destruction from UV
radiation within HII regions. Indeed, there is observational
precedence for suggesting that SN-driven shocks play a pivotal role
in the destruction of PAHs. Using ISOCAM-CVF spectra, \citet{rea02}
do not detect PAH emission within either ionic or molecular shocks
in a number of Galactic supernova remnants 3C 391, despite strong
PAH features detected in the spectra of the pre-shock region and
also in nearby regions not affected by the SN remnant. Furthermore,
the shocked emission shows no significant continuum rise at mid-IR
wavelengths. This rise is normally seen in interstellar dust
spectra, including those of quiescent regions, such as reflection
nebulae and have been attributed to the presence of very small dust
grains (VSGs) \citep{des90,mad05}. The lack of PAH emission within
shocked regions has also been noted in other Galactic SN remnants
such as G11.2-0.3 and Kes 69, based on IRAC colour ratios
\citep{rea05}. The lack of aromatic features or rising continuum in
the mid-IR spectrum of 3C 391 suggests that both the PAH carriers
and VSGs are destroyed by SN shocks. The lack of PAH in our sample
of low metallicity galaxies, plus a lack of rising mid-IR continua
for the low metallicity sample suggests very strongly that SN shocks
play a dominant role in the PAH emission deficit in these objects,
and indeed for the state of the ISM as a whole.

In order to further clarify and disentangle the contributions that
UV radiation surrounding young stars and SN shocks play in the
destruction of PAH carriers and in the evolution of the ISM in
low-metallicity galaxies, a follow-up high-resolution spectral
mapping survey of low-metallicity galaxies using IRS is critical.
Such a high spatial resolution spectroscopic survey will provide
further insight into the physics of grain and aromatic feature
destruction in high ionization starburst and shock front
environments.

\subsubsection{Lack of ISM enrichment by PAHs}

An  alternate scenario  for the  low PAH  emission in  low
metallicity galaxies  is that  due to  the extreme  youth of  these
galaxies, not enough time has elapsed for stars  below 8 M$_{\odot}$
to move off the MS to become Asymptotic Giant Branch (AGB) stars.
These evolved objects are considered to be the primary source  for
PAHs \citep{mor03}. An AGB star consists of  a degenerate  C/O core
surrounded by  a very extended convective atmosphere, from which
mass is lost via a dense and dusty outflow at rates of 10$^{-8}$ to
10$^{-4}$ M$_{\odot}$ yr$^{-1}$ and at expansion velocities of  5-30
km s$^{-1}$ \citep{mat05}.  PAH  carriers from AGB stars take
$\sim$200 Myr  to be injected  into the ISM,  when $\sim$4
M$_{\odot}$ stars evolve off the MS \citep{dwek05}, thus requiring
several generations of star formation to enrich the ISM with PAH
carriers.

However,  AGB  stars have  been  detected  even  in the  lowest
known metallicity galaxy,  I Zw  18 \citep{izo04b}, and  the age of
the AGB population ($\sim$500 Myr)  is more than long enough  for
injection of PAHs into  the ISM, and for  the ISM to become  PAH
enriched. However, modeling by  \citet{mor03} suggests that  despite
the presence  of AGB stars, a timescale of  up to 1 Gyr is required
to  enrich the ISM with PAH to the level we  see in higher
metallicity galaxies. Nevertheless the presence of AGB stars plus
the expected level of PAH enrichment after 500 Myr (from the models
of Morgan and Edmunds (2003)), indicate that we should see higher
PAH emission than has been detected in low-metallicity galaxies such
as I Zw 18. Given the high [FeII]/[NeII] ratios for low-metallicity
galaxies, it would be plausible to suggest that shocks  from SN
further depress PAH emission in  these galaxies through destruction
of the PAH carriers. Such a scenario may help to explain the lack of
PAH emission even in galaxies with advanced AGB populations. The
lack of PAH emission within SN shocks such as those within the
Galactic SN remnant 3C 391 \citep{rea02} would strongly enhance the
plausibility of this scenario.

\subsection{Is there evidence of a variation in the upper mass
limit with increasing metallicity?}

From discussions of the evolution of fine structure line ratios in
Sect.~3.1 and 3.2.1, we note that the decrease in the strength of
the fine structure line ratios with higher metallicity is consistent
with the hardness of the ionizing radiation in the starburst
increasing with decreasing metallicity \citep{camp86,mad05,wu05}. It
has been suggested that the formation of massive stars is directly
related to lower metallicities within the ISM, with a larger Jeans
mass being formed due to a drop in the cooling efficiency of the
primordial gas \citep{brom02} leading to a higher ionization
environment within the starburst regions. Given the proclivity for
such low-metallicity objects to produce such massive stars, it would
be interesting to determine if the decrease in the [NeIII]/[NeII]
ratio with higher metallicity is indicative of a variation in the
upper mass limit of the initial mass function (IMF) with increasing
metallicity. Theoretical studies predict that the production of very
massive stars  in high metallicity galaxies is suppressed through
feedback from previous stellar generations \citep{moy01,rig04}. The
drop in the [NeIII]/[NeII] ratio with increasing metallicity and PAH
strength would suggest a lack of very massive stars (with initial
masses greater than 40 M$_{\odot}$) being formed and going onto
produce supernovae as the ISM evolves through feedback from earlier
generations. In support of this scenario, photoionization work by
\citet{rig04} and \citet{mar05} attribute low excitation ratio
values to high-mass, solar-metallicity starbursts forming fewer
massive stars through feedback into the ISM. However, since the
[NeIII]/[NeII] ratio is dependent on both the upper mass cutoff
(M$_{upper}$) and the age of the stellar population (e.g.
\citet{leith99}), it is difficult to decouple one effect from the
other. Further detailed modeling using stellar population synthesis
codes coupled with photoionization codes using a large number of
line diagnostics could help constrain M$_{upper}$.

\section{Conclusions}

Using archival spectral data from {\it Spitzer}, we undertook a
survey of 18 starburst  galaxies in order to determine  whether SN
shocks are responsible for the PAH deficiency in low-metallicity
galaxies. Using the emission line ratio of [FeII] at 26 $\mu$m to
[NeII] at 12.8 $\mu$m  plotted   against  the  PAH strength,  we
found   a  strong anti-correlation,  with   the [FeII]/[NeII]  ratio
decreasing  with increasing  PAH emission. Similar correlations were
found   when  comparing the [FeII]/[NeII] ratio  to   the  PAH/IR
luminosities and metallicity. Since  [FeII] emission has  been
linked primarily to SN shocks, we  attribute the high [FeII]/[NeII]
ratios in low-metallicity galaxies to enhanced SN activity, and
consider this to be  the  dominant  mechanism  for  PAH destruction,
rather  than grain destruction   in  photoionized regions
surrounding   young massive stars.   We  also   consider whether the
extreme   youth of  the low-metallicity  galaxies is responsible for
the  lack of  PAH emission.  While  the  age of the known  AGB
populations in  our low-metallicity  sample galaxies is less than
the time required for full PAH  enrichment of the ISM, we argue that
SN shocks (as evidenced by from the high [FeII]/[NeII] ratios and SN
rates) further depress the PAH emission. We conclude that while SN
shock destruction  of the  PAH carriers may not be fully responsible
for the lack of PAH emission, it remains  a  prime culprit  for  the
lack  of  PAH  emission  in  low metallicity galaxies.

\acknowledgments

This work is based on archival data obtained with the {\it Spitzer
Space Telescope}, which is operated by the Jet Propulsion
Laboratory, California Institute of Technology under a contract with
NASA. We wish to thank J. Weingartner for helpful comments and
guidance, and of course the referee for providing comments and
feedback on our original submission.

   \clearpage

   \begin{figure}
   \resizebox{\columnwidth}{!}  {\includegraphics*{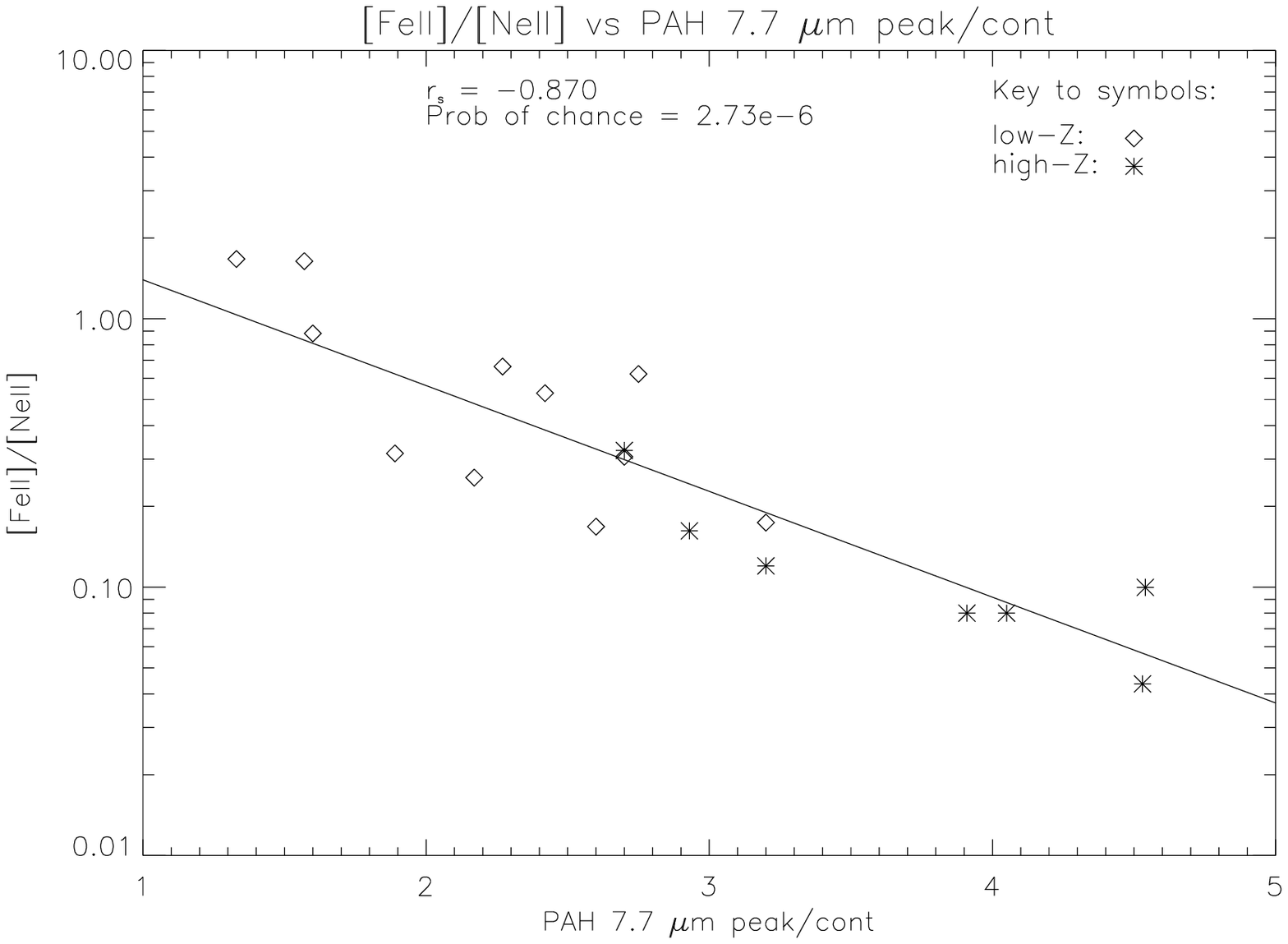}}

   \caption{Plot of the [FeII]/[NeII] ratio as a function of the PAH strength for the sample. The
    Spearman rank coefficient (r$_{s}$) of -0.870 and the very low probability of chance correlation value
    (P$_{s}$)  of 2.73 x 10$^{-6}$ confirm that a highly significant anti-correlation exists between the
    [FeII]/[NeII] ratio and the PAH strength. This strong correlation suggests that supernova-driven shocks
    are the dominant mechanism responsible for the PAH deficiency in low metallicity galaxies  - this is discussed
    further in Sect. 3.1.}

   \end{figure}

    \clearpage

   \begin{figure}
   \resizebox{\columnwidth}{!}  {\includegraphics*{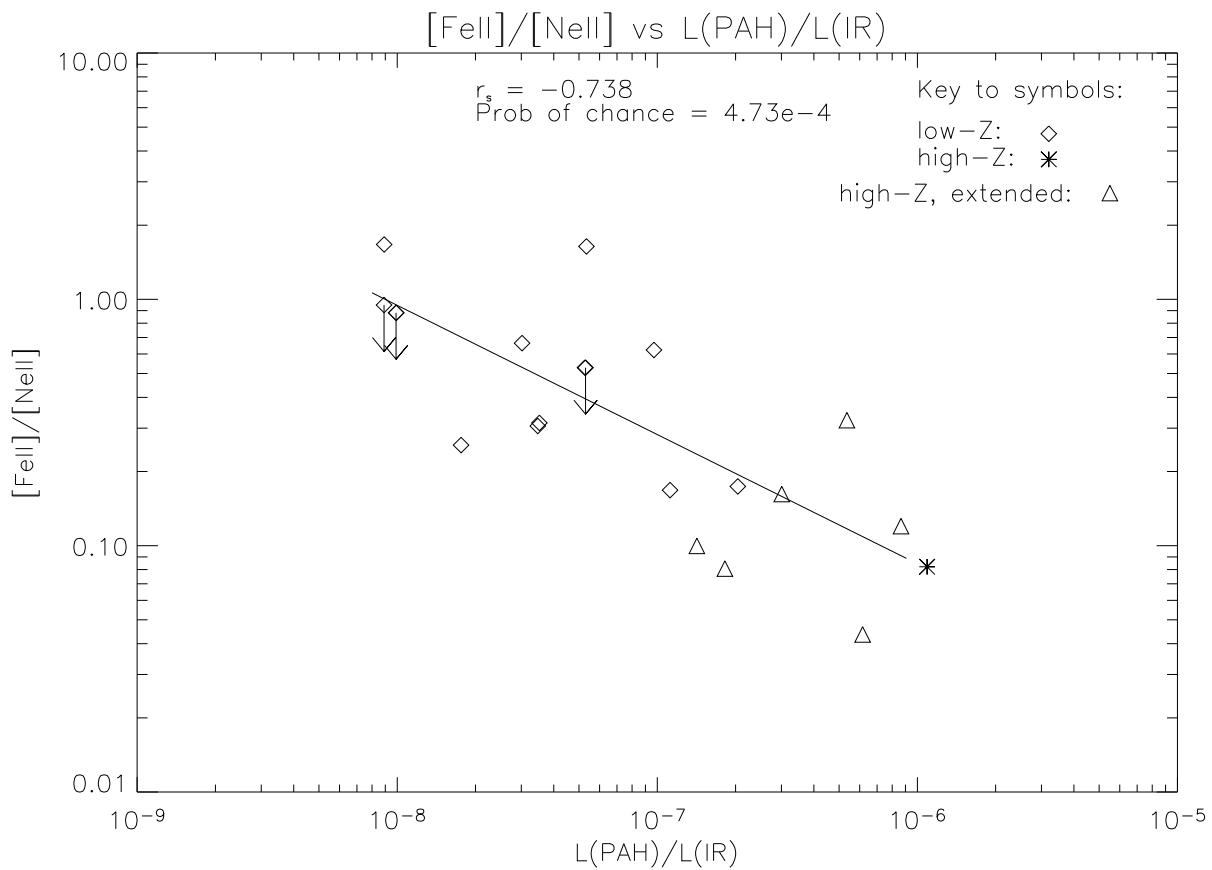}}

   \caption{Plot of the [FeII]/[NeII] ratio as a function of the PAH/IR luminosity ratio. The Spearman rank
   coefficient (r$_{s}$) of -0.738 and the very low probability of chance correlation value (P$_{s}$)  of
   4.73 x 10$^{-4}$ confirm that a  strong anti-correlation exists between the [FeII]/[NeII] ratio and the
   PAH/IR luminosity ratio. The extended high metallicity sources are indicated by the triangles.
   This correlation, along with that in Fig. 1, suggests that supernova-driven shocks
   are the dominant mechanism responsible for the PAH deficiency in low metallicity galaxies.}

   \end{figure}

 \clearpage

   \begin{figure}
   \resizebox{\columnwidth}{!}  {\includegraphics*{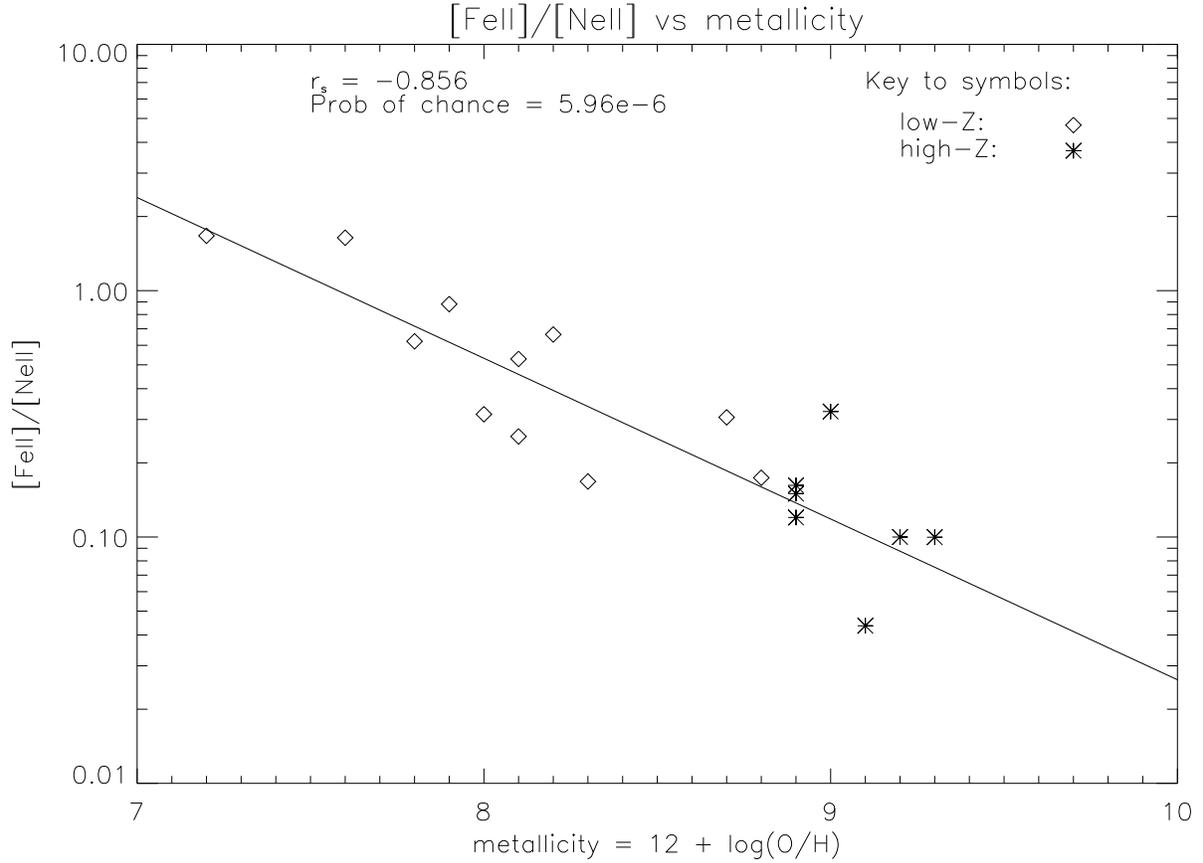}}

   \caption{Plot of the [FeII]/[NeII] ratio as a function of the galaxy metallicity. The Spearman rank coefficient
   (r$_{s}$) of -0.856 and the very low probability of chance correlation value (P$_{s}$)  of 5.96 x 10$^{-6}$
   show that a very robust anti-correlation exists between the [FeII]/[NeII] ratio and the metallicity of the galaxy.
   This strong anti-correlation, along with those in Figs. 1 and 2, would firmly suggest that supernova-driven shocks
   are the dominant mechanism responsible for the PAH deficiency in low metallicity galaxies.}

   \end{figure}

  \clearpage

  \begin{figure}
  \resizebox{\columnwidth}{!}  {\includegraphics*{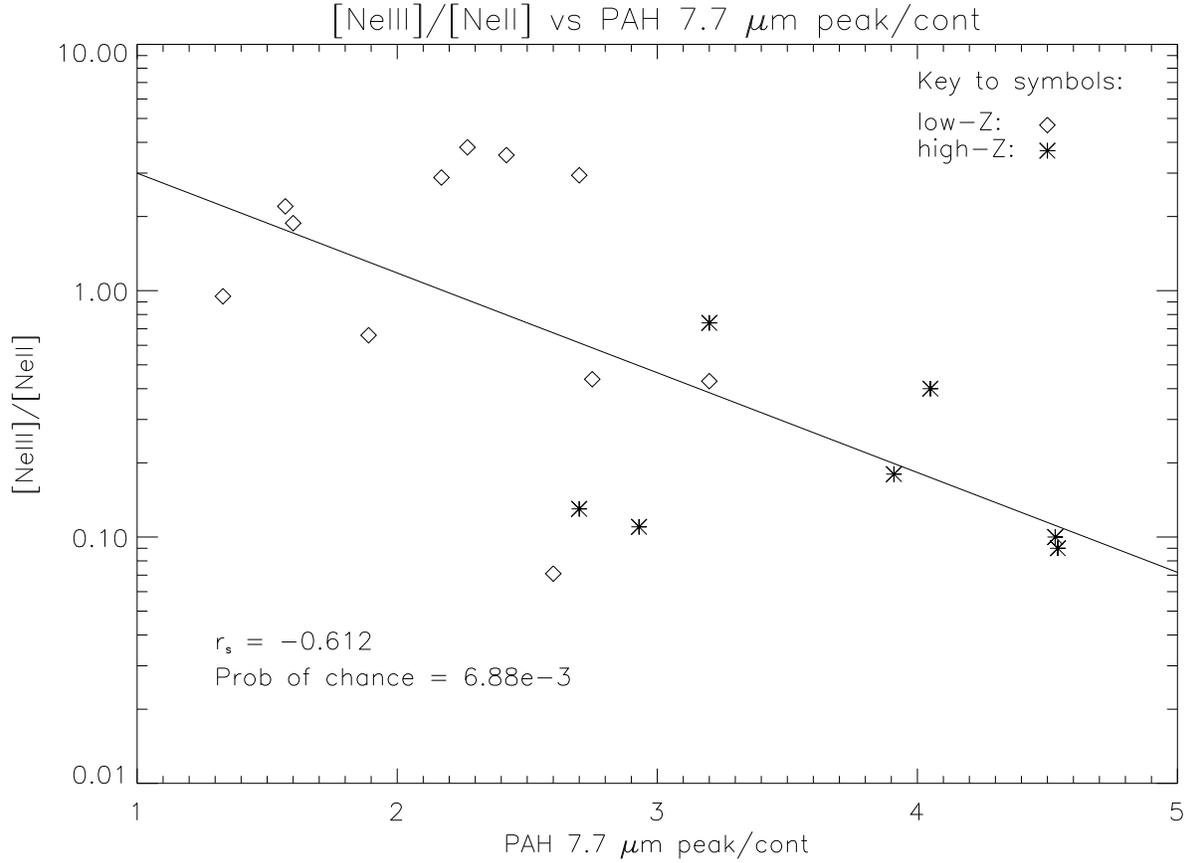}}

  \caption{Plot of the [NeIII]/[NeII] ratio as a function of the PAH strength. Although a significant correlation
(r$_{s}$ = -0.612  ; P$_{s}$= 6.88 x 10$^{-3}$), there is
considerable scatter and  the correlation is weaker  than  the
correlation displayed  in Fig.~1. This suggests that  grain
destruction  by SN-driven  shocks likely plays a  more pronounced
role than  grain destruction  within HII regions in terms of PAH
carrier destruction, given the strength of the correlations in Figs.
1-3.}

  \end{figure}

\clearpage

\begin{deluxetable}{ccccccc}
\tabletypesize{\scriptsize}

\tablecaption{Sample parameters.  \label{tbl-1}}

 \tablewidth{0pt}
 \tablehead{\colhead{Target}    &
 \colhead{R.A.}    &
 \colhead{Dec.}    &
 \colhead{Hubble type}    &
 \colhead{Distance}    &
 \colhead{Metallicity}    &
 \colhead{L(IR)}          \\
 \colhead{}   &
 \colhead{}   &
 \colhead{}   &
 \colhead{}   &
 \colhead{(Mpc)}   &
 \colhead{($\it Z_{\odot}$)}   &
 \colhead{(L$_{\odot}$)} \\
 \colhead{(1)}   &
 \colhead{(2)}   &
 \colhead{(3)}   &
 \colhead{(4)}   &
 \colhead{(5)}   &
 \colhead{(6)}   &
 \colhead{(7)}}

  \startdata

  I Zw 18   & 09\(^{h}\)   34\(^{m}\)  02.0\(^{s}\)& 55$^{\circ}$   14$'$  28$''$ & cI BCD               & 10.6 & 7.2\tablenotemark{a} & $\leq$ 3.2 x 10$^{8}$ \\
  Tol 65    & 12\(^{h}\)   25\(^{m}\)  46.9\(^{s}\)& -36$^{\circ}$  14$'$  01$''$ & HII                  & 38.0 & 7.6\tablenotemark{b} & $\leq$ 5.0 x 10$^{8}$  \\
  Mrk 25    & 10\(^{h}\)   03\(^{m}\)  51.8\(^{s}\)& 59$^{\circ}$   26$'$  10$''$ & E/S0 HII             & 42.1 & 7.8\tablenotemark{c} & 7.5 x 10$^{9}$ \\
  Mrk 170   & 11\(^{h}\)   26\(^{m}\)  50.4\(^{s}\)& 64$^{\circ}$   08$'$  17$''$ & Pec                  & 13.8 & 7.9\tablenotemark{d} & 2.9 x 10$^{8}$ \\
  UM 448    & 11\(^{h}\)   42\(^{m}\)  12.4\(^{s}\)& 00$^{\circ}$   20$'$  03$''$ & Sb pec;Sbrst HII     & 78.4 & 8.0\tablenotemark{a} & 7.2 x 10$^{10}$ \\
  II Zw 40  & 05\(^{h}\)   55\(^{m}\)  42.6\(^{s}\)& 03$^{\circ}$   23$'$  33$''$ & BCD/Sbc              & 11.1 & 8.1\tablenotemark{a}  & 3.5 x 10$^{9}$ \\
  Mrk 930   & 23\(^{h}\)   31\(^{m}\)  58.3\(^{s}\)& 28$^{\circ}$   56$'$  50$''$ & Sbrst                & 77.3 & 8.1\tablenotemark{e} & 2.6 x 10$^{10}$ \\
  NGC 4670  & 09\(^{h}\)   32\(^{m}\)  10.1\(^{s}\)& 21$^{\circ}$   30$'$  03$''$ & SB(s)0/a pec:  BCDG  & 15.1 & 8.2\tablenotemark{f} & 1.9 x 10$^{9}$ \\
  IC 342    & 03\(^{h}\)   46\(^{m}\)  48.5\(^{s}\)& 68$^{\circ}$   05$'$  46$''$ & SAB(rs)cd HII        & 3.0  & 8.3\tablenotemark{g} & 2.5 x 10$^{9}$ \\
  NGC 7793  & 23\(^{h}\)   58\(^{m}\)  58.9\(^{s}\)& 03$^{\circ}$   38$'$  05$''$ & Sbc                  & 3.2  & 8.7\tablenotemark{h} & 4.4 x 10$^{8}$ \\
  NGC 4194  & 12\(^{h}\)   14\(^{m}\)  09.5\(^{s}\)& 54$^{\circ}$   31$'$  37$''$ & IBm pec;BCG    HII   & 34.4 & 8.8\tablenotemark{i} & 7.7 x 10$^{9}$ \\
  NGC 253   & 00\(^{h}\)   47\(^{m}\)  33.1\(^{s}\)& -25$^{\circ}$  17$'$  18$''$ & SAB(s)c;HII    Sbrst & 3.4  & 8.9\tablenotemark{a} & 2.6 x 10$^{10}$ \\
  He 2-10   & 08\(^{h}\)   36\(^{m}\)  15.2\(^{s}\)& -26$^{\circ}$  24$'$  34$''$ & I0? pec Sbrst        & 12.3 & 8.9\tablenotemark{a} & 5.3 x 10$^{9}$ \\
  NGC 7714  & 23\(^{h}\)   36\(^{m}\)  14.1\(^{s}\)& 02$^{\circ}$   09$'$  19$''$ & SB(s)b:pec; HII      & 39.4 & 8.9\tablenotemark{a} & 5.2 x 10$^{10}$ \\
  NGC 3049  & 09\(^{h}\)   54\(^{m}\)  49.6\(^{s}\)& 09$^{\circ}$   16$'$  18$''$ & SB(rs)ab;HII Sbrst   & 21.1 & 9.0\tablenotemark{j} & 3.9 x 10$^{9}$ \\
  NGC 1482  & 03\(^{h}\)   54\(^{m}\)  38.9\(^{s}\)& -20$^{\circ}$  30$'$  09$''$ & SA0+ pec sp HII      & 27.1 & 9.1\tablenotemark{k} & 7.1 x 10$^{10}$ \\
  M 82      & 09\(^{h}\)   55\(^{m}\)  52.2\(^{s}\)& 69$^{\circ}$   40$'$  47$''$ & I0;Sbrst HII         & 2.9  & 9.2\tablenotemark{l} & 2.9 x 10$^{10}$ \\
  NGC 2903  & 09\(^{h}\)   32\(^{m}\)  10.1\(^{s}\)& 21$^{\circ}$   30$'$  03$''$ & SB(s)d HII           & 7.8  & 9.3\tablenotemark{a} & 7.9 x 10$^{9}$ \\

  \enddata
\\

{\noindent{\scriptsize{\textbf{Column explanations}:} (1) Common
source names; (2) Right ascension; (3) Declination; (4) Galaxy
morphology in terms of Hubble type; (5) Distance of object in
Megaparsecs; (6) Metallicity of galaxy in terms of solar
metallicity, where metallicity = 12 + log (O/H); (7) Infrared
luminosity of galaxy in terms of solar luminosity, where L(IR) = 1.8
x 10$^{-14}$ x (13.48f$_{12}$ + 5.16f$_{25}$ + 2.58f$_{60}$ +
f$_{100}$).}}

{\noindent{\scriptsize{\textbf{Sources for metallicity values}:}
(a): \citet{eng05}; (b) \citet{izo04a};  (c) \citet{hop02}; (d)
\citet{pet86};  (e) \citet{izo99}; (f) \citet{lisen98}; (g)
\citet{cros01}; (h) \citet{kong99}; (i) \citet{gar02};  (j)
\citet{gon02}; (k) \citet{par03}; (l) \citet{ham99}.}} \\

\end{deluxetable}

\clearpage

\begin{deluxetable}{cccccccc}
\tabletypesize{\scriptsize}

\rotate

\tablecaption{Flux, luminosity and ratio data. \label{tbl-2}}

 \tablewidth{0pt}
 \tablehead{\colhead{Target}    &
 \colhead{F([NeII])}    &
 \colhead{F([NeIII]}    &
 \colhead{[NeIII]/[NeII] ratio}    &
 \colhead{F([FeII]) }    &
 \colhead{[FeII]/[NeII] ratio }    &
 \colhead{L([PAH]) }    &
 \colhead{PAH 7.7 $\mu$m}    \\
 \colhead{}   &
 \colhead{@ 12.8 $\mu$m}   &
 \colhead{@ 15.6 $\mu$m}   &
 \colhead{15.6/12.8 $\mu$m}   &
 \colhead{@ 26.0 $\mu$m}   &
 \colhead{26.0/12.8 $\mu$m}   &
 \colhead{7.7 $\mu$m feature}   &
 \colhead{peak/continuum}  \\
 \colhead{}   &
 \colhead{\it (W cm$^{-2}$)}   &
 \colhead{\it (W cm$^{-2}$)}   &
 \colhead{}   &
 \colhead{\it (W cm$^{-2}$)}   &
 \colhead{}   &
 \colhead{\it (W)}            &
 \colhead{} \\
 \colhead{(1)}   &
 \colhead{(2)}   &
 \colhead{(3)}   &
 \colhead{(4)}   &
 \colhead{(5)}   &
 \colhead{(6)}   &
 \colhead{(7)}   &
 \colhead{(8)} }

  \startdata

  I Zw 18   & 4.32 $\pm$ 0.10 x 10$^{-22}$ & 4.02 $\pm$ 0.58 x 10$^{-22}$ & 0.95 & 7.20 $\pm$ 2.10 x 10$^{-22}$ & 1.67 & 1.1 x 10$^{27}$ & 1.33 \\
  Tol 65    & 3.29 $\pm$ 0.39 x 10$^{-22}$ & 7.23 $\pm$ 0.50 x 10$^{-22}$ & 2.20 & 5.40 $\pm$ 0.66 x 10$^{-22}$ & 1.64 & 1.0 x 10$^{28}$ & 1.57 \\
  Mrk 25    & 1.30 $\pm$ 0.04 x 10$^{-20}$ & 5.65 $\pm$ 0.19 x 10$^{-21}$ & 0.44 & 8.10 $\pm$ 1.60 x 10$^{-21}$ & 0.62 & 2.8 x 10$^{29}$ & 2.75 \\
  Mrk 170   & 7.39 $\pm$ 1.56 x 10$^{-22}$ & 1.39 $\pm$ 0.06 x 10$^{-21}$ & 1.88 & 6.53 $\pm$ 0.75 x 10$^{-22}$ & 0.88 & 1.1 x 10$^{27}$ & 1.60 \\
  UM 448    & 1.82 $\pm$ 0.05 x 10$^{-20}$ & 1.20 $\pm$ 0.02 x 10$^{-20}$ & 0.66 & 5.73 $\pm$ 1.09 x 10$^{-21}$ & 0.32 & 9.6 x 10$^{29}$ & 1.89 \\
  II Zw 40  & 5.90 $\pm$ 0.16 x 10$^{-21}$ & 1.70 $\pm$ 0.05 x 10$^{-20}$ & 2.88 & 1.51 $\pm$ 0.41 x 10$^{-21}$ & 0.26 & 2.4 x 10$^{28}$ & 2.17 \\
  Mrk 930   & 2.65 $\pm$ 0.19 x 10$^{-21}$ & 9.40 $\pm$ 0.09 x 10$^{-21}$ & 3.55 & 1.40 $\pm$ 0.32 x 10$^{-21}$ & 0.53 & 5.2 x 10$^{29}$ & 2.42 \\
  NGC 4670  & 5.90 $\pm$ 0.44 x 10$^{-21}$ & 1.70 $\pm$ 0.14 x 10$^{-20}$ & 3.82 & 3.84 $\pm$ 0.52 x 10$^{-21}$ & 0.66 & 2.2 x 10$^{28}$ & 2.27 \\
  IC 342    & 4.58 $\pm$ 0.15 x 10$^{-19}$ & 3.24 $\pm$ 0.24 x 10$^{-20}$ & 0.07 & 7.70 $\pm$ 0.03 x 10$^{-20}$ & 0.17 & 1.1 x 10$^{29}$ & 2.60 \\
  NGC 7793  & 3.30 $\pm$ 0.17 x 10$^{-21}$ & 9.70 $\pm$ 0.57 x 10$^{-22}$ & 2.94 & 1.01 $\pm$ 0.28 x 10$^{-21}$ & 0.31 & 5.9 x 10$^{27}$ & 2.70 \\
  NGC 4194  & 1.15 $\pm$ 0.05 x 10$^{-19}$ & 4.93 $\pm$ 0.08 x 10$^{-20}$ & 0.43 & 1.24 $\pm$ 0.31 x 10$^{-20}$ & 0.17 & 6.0 x 10$^{30}$ & 3.20 \\
  NGC 253   & 3.04 $\pm$ 0.07 x 10$^{-18}$ & 2.21 $\pm$ 0.22 x 10$^{-19}$ & 0.07 & 3.08 $\pm$ 0.13 x 10$^{-19}$ & 0.16 & 3.0 x 10$^{30}$ & 2.93 \\
  He 2-10   & 2.80 $\pm$ 0.04 x 10$^{-19}$ & 1.12 $\pm$ 0.04 x 10$^{-19}$ & 0.40 & 2.30 $\pm$ 0.09 x 10$^{-20}$ & 0.08 & 2.2 x 10$^{30}$ & 4.05 \\
  NGC 7714  & 1.08 $\pm$ 0.03 x 10$^{-19}$ & 7.95 $\pm$ 0.08 x 10$^{-20}$ & 0.74 & 1.30 $\pm$ 0.16 x 10$^{-20}$ & 0.12 & 9.6 x 10$^{30}$ & 3.20 \\
  NGC 3049  & 3.00 $\pm$ 0.08 x 10$^{-20}$ & 3.95 $\pm$ 0.24 x 10$^{-21}$ & 0.13 & 9.70 $\pm$ 0.18 x 10$^{-21}$ & 0.32 & 8.0 x 10$^{29}$ & 2.70 \\
  NGC 1482  & 3.90 $\pm$ 0.07 x 10$^{-19}$ & 3.20 $\pm$ 0.32 x 10$^{-20}$ & 0.10 & 1.70 $\pm$ 0.07 x 10$^{-20}$ & 0.04 & 1.7 x 10$^{31}$ & 4.53 \\
  M 82      & 4.69 $\pm$ 0.22 x 10$^{-19}$ & 8.58 $\pm$ 0.05 x 10$^{-20}$ & 0.18 & 7.10 $\pm$ 0.23 x 10$^{-20}$ & 0.08 & 2.2 x 10$^{30}$ & 3.91 \\
  NGC 2903  & 8.52 $\pm$ 0.04 x 10$^{-19}$ & 7.95 $\pm$ 0.36 x 10$^{-20}$ & 0.09 & 8.50 $\pm$ 0.06 x 10$^{-20}$ & 0.10 & 4.3 x 10$^{29}$ & 4.54 \\

  \enddata

{\textbf{Column explanations}:} (1) Common source names; (2) Flux of
the 12.8 $\mu$m [NeII] fine structure line in Watts per meter
squared; (3) Flux of the 15.6 $\mu$m [NeIII] fine structure line in
Watts per cm squared; (4) Ratio of the [NeIII] and [NeII] fine
structure lines; (5) Flux of the 26.0 $\mu$m [FeII] fine structure
line in Watts per cm squared; (6) Ratio of the [FeII] and [NeII]
fine structure lines; (7)Luminosity of the 7.7 $\mu$m PAH feature in
Watts; (8) Ratio of the flux density of the peak of the 7.7 $\mu$m
PAH feature to the flux density of the continuum at 7.7 $\mu$m.
\end{deluxetable}

\end{document}